\documentclass[a4paper]{article}
\usepackage{twocolpceurws}
\usepackage{microtype}

\usepackage[utf8]{inputenc}

\usepackage{graphicx}
\usepackage[utf8]{inputenc}
\usepackage{xcolor}
\usepackage{amsmath}
\usepackage{array}
\usepackage{url}
\usepackage{booktabs}
\usepackage{hyperref}
\usepackage{tabulary}

\makeatletter
\newcommand{\printfnsymbol}[1]{%
  \textsuperscript{\@fnsymbol{#1}}%
}
\makeatother

\title{Semantic Driven Fielded Entity Retrieval}

\date{22 October 2018}

\author{Shahrzad Naseri\thanks{equal contribution} \and Sheikh Muhammad Sarwar\printfnsymbol{1} \and James Allan}
\institution{Center for Intelligent Information Retrieval\\
College of Information and Computer Sciences\\University of Massachusetts Amherst \\ \{shnaseri,smsarwar,allan\}@cs.umass.edu}

\begin{document}
\maketitle

\begin{abstract}
A common approach for knowledge-base entity search is to consider an entity as a document with multiple fields. Models that focus on matching query terms in different fields are popular choices for searching such entity representations.
An instance of such a model is FSDM (Fielded Sequential Dependence Model). We propose to integrate field-level semantic features into FSDM. We use FSDM to retrieve a pool of documents, and then to use semantic field-level features to re-rank those documents. We propose to represent queries as bags of terms as well as bags of entities, and eventually, use their dense vector representation to compute semantic features based on query document similarity. Our proposed re-ranking approach achieves significant improvement in entity retrieval on the DBpedia-Entity (v2) dataset over existing FSDM model. Specifically, for all queries we achieve 2.5\% and 1.2\% significant improvement in NDCG@10 and NDCG@100, respectively.   
\end{abstract}

      
\section{Introduction}
In recent years, web search engines are moving toward answering users' query with a more focused response. Examples include entity cards as well as lists of named entities such as people, organizations, and locations as the answers or query suggestions. Studies over Bing~\cite{guo2009named} and Yahoo~\cite{pound2010ad} web search queries has shown that over $70\%$ and $50\%$ of query logs are related to entities, respectively.  

The core underlying most methods that provide such focused responses is collections of Knowledge Bases (KB). Knowledge bases provide a unified view of entities and the relationships between them. Knowledge bases such as DBpedia\footnote{\url{http://dbpedia.org}}, YAGO\footnote{\url{http://www.mpi-inf.mpg.de/yago-naga/yago/}}, and Freebase\footnote{\url{http://freebase.org}} store entities information in a subject-predicate-object format, which is called Resource Description Framework (RDF) triple. Structured representation of entities available in KB made them attractive collections for entity search against natural language queries. In order to answer a users' query from the knowledge bases, the task of entity retrieval is defined as returning a ranked list of relevant entity articles to respond users' query.

Previous works represented a knowledge-base entity as a structured document by grouping RDFs into fields~\cite{balog2013test, zhiltsov2015fielded} or tree structure~\cite{lu2015entity}. For example, Zhiltsov et al.~\cite{zhiltsov2015fielded} define five fields such as names, attributes, categories, similar entity names, and related entities to represent an entity. They proposed Fielded Sequential Dependence Model (FSDM) and showed that term dependence is an important aspect for entity search. However, their work did not consider semantic matching of terms and documents which has become a popular choice for ad-hoc retrieval. 

Capturing the semantic similarity between vocabulary terms and pieces of text is a long-standing problem  in Information Retrieval (IR). Different methods have been proposed in this regard and one prevalent as well as recent choice among them is \textit{word embedding}. Word embedding encodes the semantic information associated with a word by exploiting word co-occurrence information. Word2Vec~\cite{mikolov2013distributed} and Glove~\cite{pennington2014glove} are such methods which learns a low-dimensional vector using nueral networks and matrix factorization, respectively. We propose a method for entity retrieval that computes semantic match of query and each field of a document using embeddings for words and entities present in them. 

We do our experiments on the DBpedia-Entity (v2) benchmark dataset \cite{hasibi2017dbpedia} and use the train/test split provided by them to train a model that combines FSDM score and semantic features. DBpedia is often referred as the ``database version of Wikipedia'' and it is a community effort to extract structured information from Wikipedia. We demonstrate that significant gain can be achieved for similar entity search and natural language queries as well as all queries by incorporating semantic features. All resources, including a sample of the corpus we used to learn entity embeddings, source files for our model, runs and their evaluation results are made publicly available
at \href{https://tinyurl.com/sem-fielded-entity-retrieval}{https://tinyurl.com/sem-fielded-entity-retrieval}.

The rest of this work is organized in the following manner: We provide some background on entity retrieval in section~\ref{sec:related_work}. In section~\ref{sec:approach} we discuss the formulation of our approach. Finally, we empirically validate our approach in section~\ref{sec:results} and conclude in section~\ref{sec:conclusion}.

\section{Related Work}\label{sec:related_work}

Guo et al.~\cite{guo2009named} and  Pound et al.~\cite{pound2010ad} show that over 70\% and 50\% of query logs of Bing and Yahoo, respectively, address entities. Motivated by that situation, an entity retrieval system returns ranked list of entities from a knowledge base to answer a user query.  Various benchmarking campaigns focused on this task including INEX Entity Ranking~\cite{demartini2009overview}, INEX Linked Data Track~\cite{wang2012overview}, the TREC Entity track~\cite{balog2010overview, balog2012vries, serdyukov2009delft}, the Semantic Search Challenge~\cite{blanco2011entity, halpin2010evaluating}, and the Question Answering over Linked Data (QALD) challenge series~\cite{lopez2013evaluating}. DBpedia-Entity (v2), the dataset that is used by this shared task, gathers the queries from all of these previous challenges. 

Existing methods take advantage of the fact that entities have rich fielded information and propose a variety of fielded retrieval methods such as BM25F~\cite{perez2010using, itakura2010framework, robertson2004simple} and FSDM~\cite{zhiltsov2015fielded}. In FSDM, different fields of an entity are categorized into five final fields: names, attributes, categories, related entity names, and similar entity names. FSDM incorporates term dependency based on ordered and unordered n-grams. Chen et al.~\cite{chen2016empirical} investigate learning to rank model on entity search which incorporates different features such as the FSDM score, BM25 score, etc.

There is substantial work in ad-hoc document retrieval that tries to take advantage of embeddings to improve retrieval effectiveness. Recently, Xiong et al.~\cite{xiong2017word} described a method which presents documents and queries in both text and entity space, thus leveraging entity embeddings. However, such deep models need significant amounts of data to be effective. For this task, since the provided dataset is small, our model is more readily applicable. 

Entity embeddings are also used in other tasks such as question answering~\cite{bordes2014open}, academic search~\cite{xiong2017explicit}, entity disambiguation~\cite{zwicklbauer2016robust}, and for knowledge graph completion~\cite{yang2014embedding,lin2015learning}. The TREC-CAR (Complex Answer retrieval) task provides a large dataset on a large collection of knowledge articles from Wikipedia which present an opportunity for incorporating deep models in the task of entity retrieval. TREC-CAR shows that the RDF2Vec~\cite{ristoski2016rdf2vec} is not as effective as the BM25 model in the paragraph ranking task~\cite{nanni2017benchmark}.

\section{Retrieval and re-ranking Approach}\label{sec:approach}

Our retrieval approach consists of two stages: we first create a pool of $n$ documents using FSDM \cite{zhiltsov2015fielded}, and then we re-rank them using term and entity semantic features. Zhiltsov et al. proposed considering entities as documents with five different fields and used FSDM to retrieve entities \cite{zhiltsov2015fielded}. In addition to the original five fields, another field \emph{text} containing natural language description of an entity is incorporated in our setting. 

Apart from using the top-n documents retrieved using FSDM, we use their scores in linear combination with our semantic similarity scores. We normalize the FSDM score using min-max normalization and use the result as a single feature or score in our approach. We compute two different types of similarity scores or semantic features based on two different query representations. This gives us two groups of semantic features that we linearly combine with the normalized FSDM score. We refer to the first group as ``term semantics'' and the second group as ``entity semantics''. For computing the entity semantics similarity score, we learned our own entity embedding vectors as described in Section~\ref{sec:entity2vec-learning}, however, we used the pre-trained Glove word embeddings for term semantics.   

\paragraph{Term Semantics} 
We compute the query embedding $\vec{q_t}$ using the average of the embedding of the query terms. For each field $f_i$ of a document we also use the average of the word embedding of its terms to compute the representation $\vec{d_{f_i}}$ for that specific field. Then the score of that field $f$ of a document is computed using $\cos{\vec{q_t}, \vec{d_{f_i}}}$. Finally, all the field scores are aggregated using a linear combination of the scores from each field using the following equation: 

\begin{equation*}
    \label{eq:term}
    Score_{t} = \sum_i^k \lambda^t_i \times \cos{\vec{q_t}, \vec{d_{f_i}}}
\end{equation*}
where $Score_{t}$ represents term semantics. 

\paragraph{Entity Semantics}
In this approach, we represent the query as bag-of-entities. The semantic representation $\vec{q_e}$ of a query is computed as the average of the embedding of the entities present in the query. We compute the document representation in the same way as mentioned in the previous paragraph but using entities rather than terms. The query and document representations are used to compute entity semantics using the following equation:  

\begin{equation*}
    \label{eq:entity}
    Score_{e} = \sum_i^k \lambda^e_i \times \cos{\vec{q_e}, \vec{d_{f_i}}}
\end{equation*}
where $Score_{e}$ represents entity semantics. 

\paragraph{Document Scoring}
The score of a document is computed using Equation~\ref{eq:doc_score}. It combines the term semantics, entity semantics, and normalized FSDM score. As we have six fields, we need to learn six parameters for term semantics, six for entity semantics, and one for FSDM. We learn these parameters using the Coordinate Ascent method for combining linear features proposed by Metzler et al.~\cite{metzler2007linear}. 
\begin{equation}
    \label{eq:doc_score}
    Score_D = Score_t + Score_e + \lambda \times Score_{FSDM}
\end{equation}

\subsection{Learning Entity Embeddings}\label{sec:entity2vec-learning}
Following the approach of Ni et al.~\cite{ni2016semantic}, we learned embedding vectors for entities based on the Skip-gram~\cite{mikolov2013distributed} model. To this end, we replace the hyperlinks in the Wikipedia pages (that are links to other Wikipedia pages, i.e., entities) by a placeholder representing the entity. In this case, the hyperlink mentions (i.e. phrases) will be presented as a single ``term'' and the embedding of the entity (term) can be learned using Skip-gram model. 

The following is an excerpt from Wikipedia in which entities are marked as italics:
\begin{quote}
    Albert Einstein was a German-born theoretical physicist who developed the \textit{theory of relativity}, one of the two pillars of modern physics (alongside \textit{quantum mechanics}). He is best known to the general public for his \textit{mass–energy equivalence} formula $E = mc^2$ which has been dubbed "the world's most famous equation".
\end{quote}
The excerpt will be changed to the following text in which hyperlinks (entities) are replaced by underscored of the title of the linked pages. 
\begin{quote}
    Albert Einstein was a German-born theoretical physicist who developed the \texttt{Theory\_of\_relativity}, one of the two pillars of modern physics (alongside \texttt{Quantum\_mechanics}). He is best known to the general public for his \texttt{Mass–energy\_equivalence} formula $E = mc^2$ which has been dubbed "the world's most famous equation".
\end{quote}

\section{Experimental Setup}\label{sec:experimental_setup}
In this section, we introduce datasets (beyond DBpedia-Entity (v2)) that we used in our model. We also present our data processing approaches and hyperparameter settings. 

\subsection{Data Set}
Our experiments are done using the dataset provided by the task, DBpedia-Entity (v2)~\cite{hasibi2017dbpedia}. We have used the same train/test split provided by them. We used the first 10 queries from training data to form our validation set. For embedding terms in queries and documents we used GloVe~\cite{pennington2014glove} pre-trained word embeddings. The word embeddings were originally learned from a 6 billion token collection (the Wikipedia dump 2014 plus the Gigawords 5). The entity embeddings are learned from the DBpedia 2016-10 full article Wikipedia pages dump. 
\subsection{Data Processing}
We used the FSDM run in the DBpedia-Entity (v2) collection as the baseline method. We also consider the documents retrieved in that run as our initial document pool and re-ranked them using semantic features and FSDM score. For annotating entities in the query, we used the TagMe~\cite{ferragina2012fast} mention detection tool. To learn the entity embeddings, we used the Word2Vec implementation in  gensim~\cite{rehurek_lrec}. Using the approach illustrated in Section~\ref{sec:entity2vec-learning}, we learned embeddings of 3.0M entities out of 4.8M entities available in Wikipedia.

\subsection{Hyperparameter settings}\label{sec:hyperparam}
For learning the entity embedding vectors with 200 dimensions using Skip-gram model, we used the following hyperparameters: window-size=10, sub-sampling=$1\epsilon-3$, cutoff min-count=0. 

To learn the weights in our model, we used the coordinate ascent (CA) algorithm~\cite{metzler2007linear} to directly optimize  NDCG@10. We start with random weights for all the features and use maximum 25 iterations with 2 restarts. We used the implementation of CA available at \cite{Chengithub}. 

\section{Experimental Results Discussion}\label{sec:results}
 
Table \ref{table:result} shows the result of incorporating semantic information with scores of our baseline FSDM model. Term semantics refers to the similarity scores obtained for different fields of a document by considering the query as a term vector, while entity semantics consider the query as a bag of entities. We report the results based on different query groups in DBpedia-Entity v2 dataset. For the convenience of discussing our results, we provide one example for each type of query in Table \ref{tab:query_type}. 

Our results show that we achieve improvement by incorporating semantics (term semantics, entity semantics, or a combination of both) over all the query types. Incorporating entity semantic achieves the highest improvement in all query types except ListSearch queries. Note that for including entity information we only consider the query as a bag of entities. As a result of that choice, the converted list query ``Professional sports teams in Philadelphia'' would have two entities: \emph{Professional sports teams} and \emph{Philadelphia}. However, a \emph{ListSearch} query is comprises three components: the target entity which is the entity to be retrieved, the source entity (\emph{Philadelphia}), and the terms (\emph{sports}, \emph{teams}, \emph{Professional}) that specify the relation between the target entity and the source entity. Our bag-of-entities query merges the terms that specify the relations between entities and is thus not helpful for that class of queries. As a consequence, we can see that incorporation of term semantics results in better performance compared to entity semantics in list search. Query term merging in this case might have been helpful if we have considered category or type embedding. Our work is more focused towards entity embedding, and we leave incorporating type embedding as future work. 

Our approach yields the maximum (and significant) improvement for QALD query type. Including entity semantics resulted in 6.7\% and 3.7\% improvement over the FSDM baseline in term of NDCG@10 and NDCG@100, respectively. All these indicate that both term and entity semantics gives valuable gains in re-ranking. 

Finally, we see significant improvement when we consider all the queries together. In this case, including both term and entity semantics resulted in the best NDCG@10. This is a statistically significant improvement over the baseline FSDM model. We also achieve significant improvement in NDCG@100 by incorporating entity semantics. However, in some cases incorporation of both term and entity semantics do not result in better performance compared to individually including them because of the increase in the number of features and lack of training data.        

\begin{table}[]
    \caption{Query types in DBpedia-Entity (v2) and their examples 
    \cite{hasibi2017dbpedia}}
    \centering
    \begin{tabulary}{\linewidth}{p{6em}|L}
    \hline
    Query Type & Example \\
    \hline \hline
    INEX-LD & \emph{Electronic music geners} \\
    ListSearch & \emph{Professional sports teams in Philadelphia} \\
    QALD-2 & \emph{Who is the mayor of Berlin?} \\
    SemSearchES & \emph{Brooklyn Bridge} \\
    \hline
    \end{tabulary}
    \label{tab:query_type}
\end{table}

\begin{table} 
\caption{Overall accuracy on each query group as well as all queries. The semantic and FSDM score are linearly combined using Coordinate Ascent algorithm.  $\dagger$ indicates significant (p < 0.05) improvement over the FSDM baseline measured by the Student’s paired t-test.}
\label{table:result}
\begin{tabulary}{\linewidth}{L|cc}
\toprule 
& \multicolumn{2}{c}{INEX\_LD}\\
Methods & NDCG@10 & NDCG@100 \\
\hline
FSDM  & 0.4214 & 0.5043 \\
FSDM + Entity Semantics & \textbf{0.4335} & \textbf{0.5119} \\
FSDM + Term Semantics& 0.4224 & 0.5015\\
FSDM + Entity \& Term Semantics & 0.4291 & 0.5047\\
\hline
& \multicolumn{2}{c}{ListSearch}\\
Methods & NDCG@10 & NDCG@100 \\
\hline
FSDM & 0.4196 & 0.4952\\
FSDM + Entity Semantics & 0.4247 & 0.4899\\
FSDM + Term Semantics & \textbf{0.4272} &  \textbf{0.5004}\\
FSDM + Entity \& Term Semantics & 0.4242 &  0.4841\\
\hline
& \multicolumn{2}{c}{QALD2}\\
Methods & NDCG@10 & NDCG@100 \\
\hline
FSDM & 0.3401 & 0.4358\\
FSDM + Entity Semantics & \textbf{0.3628}$\dagger$ & \textbf{0.4521}$\dagger$\\
FSDM + Term Semantics & 0.3390 & 0.4291\\
FSDM + Entity \& Term Semantics & 0.3448 & 0.4330\\
\hline
& \multicolumn{2}{c}{SemSearchES}\\
Methods & NDCG@10 & NDCG@100 \\
\hline
FSDM & 0.6521 & 0.7220\\
FSDM + Entity Semantics & \textbf{0.6586} & \textbf{0.7281}\\
FSDM + Term Semantics & 0.6500 &  0.7173\\
FSDM + Entity \& Term Semantics & 0.6583 & 0.7273\\
\hline
& \multicolumn{2}{c}{all\_queries}\\
Methods & NDCG@10 & NDCG@100 \\
\hline
FSDM & 0.4524 & 0.5342\\
FSDM + Entity Semantics & 0.4619$\dagger$ &  \textbf{0.5408}$\dagger$\\
FSDM + Term Semantics& 0.4617$\dagger$ &   0.5387$\dagger$\\
FSDM + Entity \& Term Semantics& \textbf{0.4639}$\dagger$   & 0.5369\\
\bottomrule
\end{tabulary}
\end{table}

\section{Conclusion and Future Works}\label{sec:conclusion}

In this study, we improve the accuracy of entity ranking by incorporating the similarity gained by comparing the query with each field of an entity document both in term and entity space. We demonstrate the efficiency of this model on a comprehensive benchmark dataset in comparison with the original FSDM model. In our experiments, we achieve statistically significant improvements over all of the queries. In order to increase the capacity of our model, we intend to learn separate vector embeddings for each field based on the content. For example, type embedding for the category field . Furthermore, we plan to adopt pairwise Learning to Rank (LTR) to determine feature weights. Moreover, by getting inspiration from the original FSDM paper which incorporates term dependencies, we hope to explore deep neural models such as RNN and LSTM in order to capture term sequence. 



\section*{Acknowledgement}
This work was supported in part by the Center for Intelligent Information Retrieval and in part by NSF grant \#IIS-1617408. Any opinions, findings and conclusions or recommendations expressed in this material are those of the authors and do not necessarily reflect those of the sponsors. 

\bibliographystyle{abbrv}
\bibliography{bibliography}

\end{document}